\documentclass[fleqn,10pt]{wlscirep}
\usepackage[utf8]{inputenc}
\usepackage[T1]{fontenc}
\usepackage{bm}

\title{Numerical solution of nonlinear Schrödinger equation by a hybrid pseudospectral-variational quantum algorithm}

\author[1,*]{Nikolas Köcher}
\author[2]{Hendrik Rose}
\author[3]{Sachin S. Bharadwaj}
\author[4,3]{Jörg Schumacher}
\author[1,2,5]{Stefan Schumacher}
\affil[1]{Department of Physics and Center for Optoelectronics and Photonics Paderborn (CeOPP), Paderborn University, D-33098 Paderborn, Germany}
\affil[2]{Institute for Photonic Quantum Systems (PhoQS), Paderborn University, D-33098 Paderborn, Germany}
\affil[3]{Tandon School of Engineering, New York University, New York City, NY 11201, USA}
\affil[4]{Institute of Thermodynamics and Fluid Mechanics, Technische Universität Ilmenau, D-98684 Ilmenau, Germany}
\affil[5]{Wyant College of Optical Sciences, University of Arizona, Tucson, AZ 85721, USA}

\affil[*]{nikolas.koecher@uni-paderborn.de}

\begin{abstract}
The time-dependent one-dimensional nonlinear Schrödinger equation (NLSE) is solved numerically by a hybrid pseudospectral-variational quantum algorithm that connects a pseudospectral step for the Hamiltonian term with a variational step for the nonlinear term. The Hamiltonian term is treated as an integrating factor by forward and backward Fourier transforms, which are here carried out classically. This split allows us to avoid higher-order time integration schemes, to apply a first-order explicit time stepping for the remaining nonlinear NLSE term in a variational algorithm block, and thus to avoid numerical instabilities. We demonstrate that the analytical solution is reproduced with a small root mean square error for a long time interval over which a nonlinear soliton propagates significantly forward in space while keeping its shape. We analyze the accuracy and complexity of the quantum algorithm, the expressibility of the ansatz circuit and compare it with classical approaches. Furthermore, we investigate the influence of algorithm parameters on the accuracy of the results, including the temporal step width and the depth of the quantum circuit.
\end{abstract}
\begin{document}

\flushbottom
\maketitle
\thispagestyle{empty}

\section*{Introduction\label{sec:introduction}}
The question on the application of quantum computing methods for the solution of linear and nonlinear ordinary or partial differential equations has received substantial interest in the past years \cite{Montanaro2016,Deutsch2020,Cerezo2021,Succi2023,Tennie2025,bharadwaj2025}. It is yet an open point if the specific and unique properties of quantum algorithms in comparison to classical ones lead to faster and more efficient numerical solution methods leaving aside the technological hurdles of present noisy intermediate scale quantum (NISQ) devices \cite{Preskill2018}. This comprises their encoding capabilities, which grow exponentially with the qubit number, and the unique parallelism and correlations due to the entanglement of multiple qubits. The question has been approached from different methodological directions and problem tasks. Fundamental nonlinear partial differential equations such as the nonlinear Schrödinger or Burgers equations appear in many applications ranging from nonlinear optics \cite{Fibich2015, Lueders2023}, Bose-Einstein condensation \cite{Bronski2001}, oceanography \cite{Dudley2019} via fluid turbulence \cite{Bec2007} to astrophysical jets \cite{Kofman1992}. The one-dimensional time-dependent nonlinear Schrödinger equation (NLSE) is thus often considered as a benchmark case for the application of quantum algorithms to solve a partial differential equation.

Quantum algorithms for the solution of one-dimensional (nonlinear) equations comprise (i) a combination of quantum linear systems algorithms (QLSA) with homotopy methods \cite{Xue2021,bharadwaj2024(2)}, (ii) linearization of nonlinear problems \cite{Liu2021,Itani2022,gonzalez2024} by Carleman embedding \cite{Carleman1932}, and (iii) quantum feature map encodings of nonlinearities \cite{Kyriienko2021} as in kernel methods of machine learning \cite{Schuld2019}. Some of the methods that we just mentioned, start from or go back to a linear algorithm \cite{bharadwaj2023hybrid,bharadwaj2024(1)}, such that the problem can be mapped on a quantum computer which follows the laws of {\em linear} quantum mechanics. Variational quantum algorithms (VQA) \cite{Lubasch2020,Jaksch2023,pool2024nonlinear} are inspired by variational quantum eigenvalue solvers. The latter class of variational algorithms aim to find a ground state of a molecule or a many-particle quantum system by minimizing an energy functional \cite{Peruzzo2014} or perform an initialization of wave packets for the nonadiabatic molecular quantum dynamics in chemical reactions \cite{Ollitrault2020} . This is similar to the classical Rayleigh-Ritz method for the approximate determination of eigenvalues \cite{MacDonald1933}. The class of variational methods provides the motivation for the present study as it can be implemented for the solution of nonlinear dynamics, such as a nonlinear Schrödinger equation (NLSE).   

\begin{figure}[t]
	\centering
	\includegraphics[width = 0.9\columnwidth]{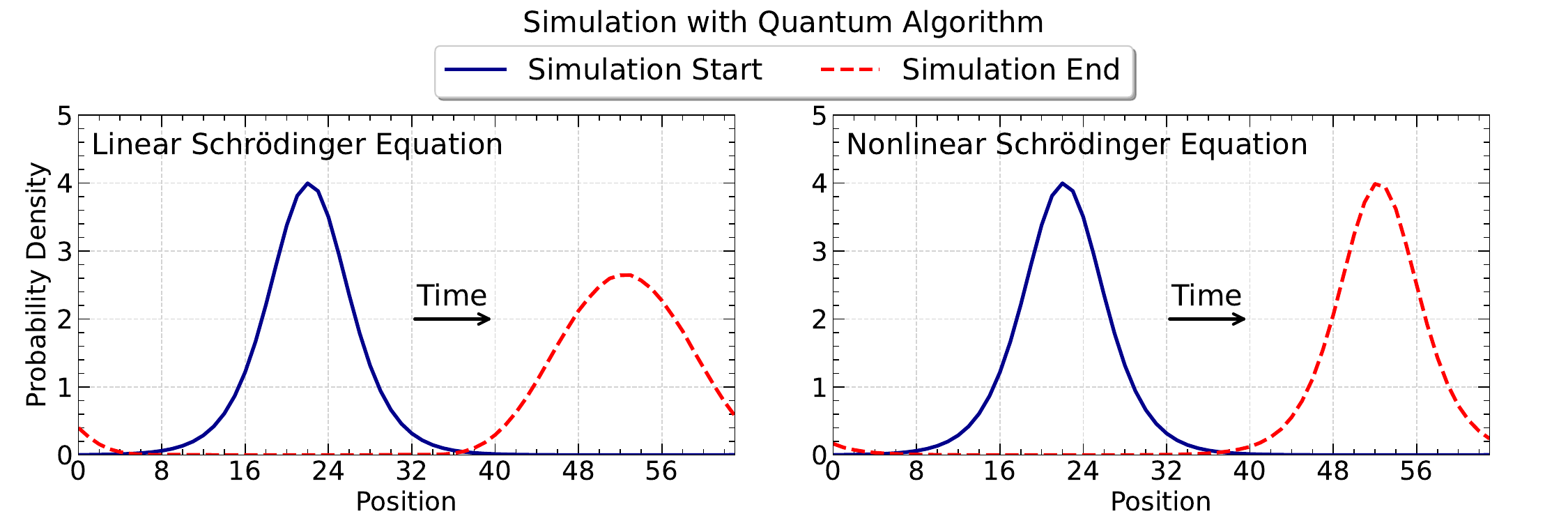}
	\caption{Illustrative simulation result obtained with the quantum algorithm presented in this paper. Left: time evolution of a wave packet in the linear Schrödinger equation; the quadratic dispersion leads to broadening with increasing propagation distance. Right: time evolution of a bright soliton in the nonlinear Schrödinger equation; the initial wave packet maintains its shape.}
	\label{fig:toc}
\end{figure}

Here, we present a numerical solution method of the one-dimensional NLSE which applies a pseudospectral-variational quantum algorithm. This algorithm combines a pseudospectral split-step method for the linear part of the NLSE and a variational algorithm for the nonlinear part. The operator splitting neglects terms of order $\mathcal{O}(\Delta t^2)$ and is therefore first-order accurate. Variational algorithms typically rely so far on first-order schemes for the time advancement \cite{Lubasch2020,Ingelmann2024,sarma2024}. In case of the NLSE this causes numerical instabilities such that higher-order time stepping methods have to be applied classically. The present split method overcomes this stability problem. Our framework resembles approaches, which have been applied in past quantum simulations of the linear Schrödinger equation, where kinetic and potential operators of the Hamiltonian have been treated separately \cite{Benenti2008}, or in hybrid optimization procedures to prepare initial states \cite{Ollitrault2020} or to obtain parametric circuits in quantum optimal control problems \cite{Huang2023}.

Furthermore, we investigate the dependence of the results on the degree of entanglement of the parametric quantum circuit. This degree of entanglement depends on the depth of the quantum circuit which is the dependence to be investigated. It is shown that the algorithm can solve the equation for a longer period of at least up to 100 integration time steps, which is an exceptionally long integration period for a quantum algorithm. 

The one-dimensional NLSE, which is also known as the Gross-Pitaevskii equation in the context of Bose-Einstein condensation, for the complex wave function $\Psi(x,t)$ is given in dimensionless form by 
\begin{equation}
i \frac{\partial \Psi}{\partial t}= -\frac{1}{2} \frac{\partial^2\Psi}{\partial x^2}+V\Psi-s |\Psi|^2 \Psi\,,\quad \Psi(x,0)=\Psi_0(x)\,,
\label{nlse1}
\end{equation}
where the nonlinear coupling constant $s$ describes the strength of the nonlinearity. The potential energy operator is $V$. The case $s>0$ gives rise to a focusing nonlinearity and bright soliton solutions; the case $s<0$ leads to defocusing nonlinearities and dark solitons. The NLSE is integrable in one dimension. For the case of $V=0$, there exist several analytical solutions of \eqref{nlse1}. For $x\in \mathbb{R}$, $s=1$, and the initial condition 
\begin{equation}
\Psi_0(x)=a\, \mbox{sech}(a(x-x_0)) e^{iv(x-x_0)}\,, 
\label{nlse2}
\end{equation}
one gets the following analytical solution
\begin{equation}
\Psi(x,t)=a\, \mbox{sech}(a(x-x_0-vt)) e^{iv(x-x_0)+\frac{i}{2}(a^2-v^2)t}\,,
\label{ana}
\end{equation}
with constants $a>0$, $v>0$, and $x_0\in \mathbb{R}$. This solution corresponds to soliton propagation. We can furthermore construct a solution $\Psi_p(x,t)$ that satisfies periodic boundary conditions on a finite domain of length $L$, 
\begin{align}
\Psi_p(x,t) = \max(\left\{  \Psi(x+k L) | k\in \mathbb{Z}\right\}). \label{ana_p}
\end{align}
This analytical solution will be used as the test case for the present algorithm. Throughout this manuscript $V=0$. Figure~\ref{fig:toc} shows an illustrative simulation result obtained with the algorithm presented in this work. 

Variational quantum algorithms -- hybrid quantum-classical algorithms -- have been used to solve linear and nonlinear partial differential equations in the past years. This includes linear one-dimensional advection-diffusion equations for simple transport problems \cite{Demirdjian2022,Leong2022,Leong2023,Ingelmann2024}, including Dirichlet and Neumann boundary conditions \cite{Over2025} and heat equations in one and two dimensions \cite{Guseynov2023,Liu2023}. Nonlinear problems, which were addressed by VQA, include finding the ground state of the steady one-dimensional NLSE \cite{Lubasch2020} and the one-dimensional Burgers equation for the nonlinear steepening of a sine wave \cite{Lubasch2020,Pool2022,pool2024nonlinear}. In refs. \cite{Pool2022,pool2024nonlinear} an alternative Feynman-Kitaev algorithm is used which orders spatial and temporal qubits in one register and thus avoids the time stepping. However, this enhances the number of qubits in the required quantum register and thus limits an application for current NISQ devices to a few time steps only. Our detailed analysis presents a further use case of a time-dependent nonlinear partial differential equation solved with this class of quantum algorithms. 

\section*{Results}
\begin{figure}[t]
	\centering
	\includegraphics[width = 0.9\columnwidth]{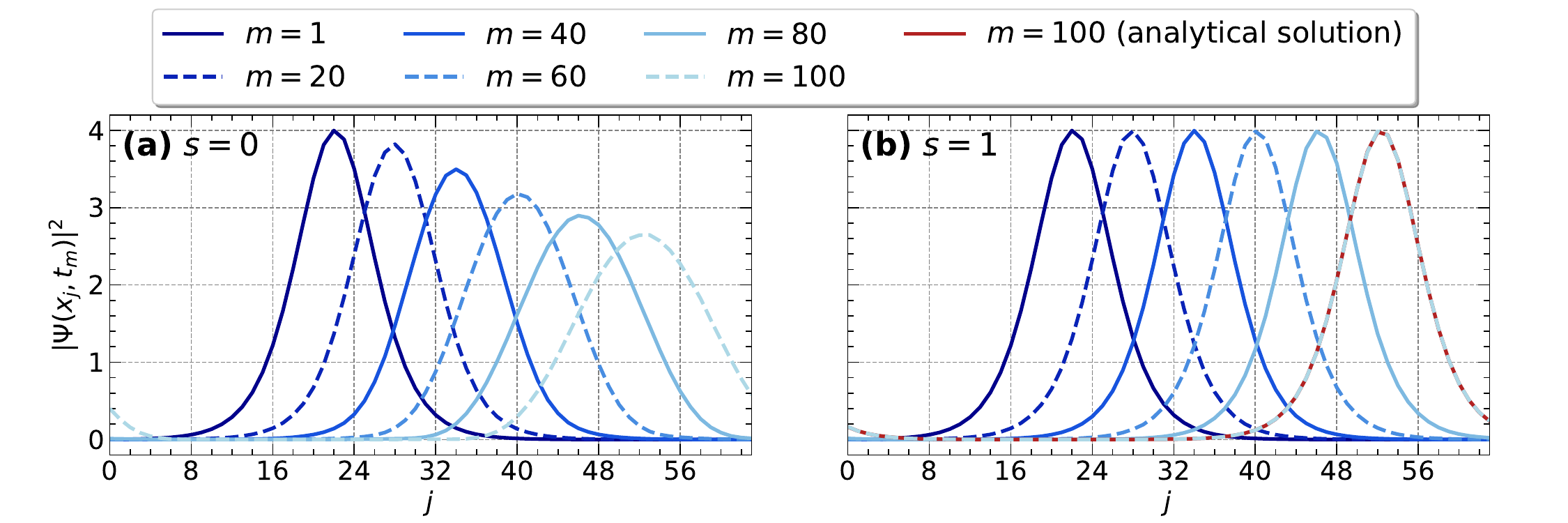}
	\caption{Simulation results obtained from the quantum algorithm introduced in the methods section are shown for selected time steps $t_m$ for (a) the linear Schrödinger equation $s=0$ and (b) the NLSE for $s=1$. The results are obtained for a qubit number of $n=6$ with the statevector simulator. The discrete output time steps are denoted as $m$ in the legend. Line styles given are for both panels. For $s=1$, the analytical solution is shown for the last time step.}
	\label{fig:lin_and_nonlin}
\end{figure}

This section is composed of three parts. First, we will demonstrate the simulation results with the hybrid algorithm and discuss its suitability and accuracy. Secondly, we will investigate the importance of algorithm parameters, such as the width of the integration timestep and the depth of the parametric quantum circuit for the performance. These points are connected to the expressibility of the quantum circuit. Finally, we will discuss the complexity of the hybrid algorithm.

\subsection*{Simulation Results and Accuracy Analysis} \label{sec:results_a}

The hybrid pseudospectral-variational quantum algorithm proceeds in two steps, an implicit first step for the linear Laplacian term in \eqref{nlse1} and an explicit step for the nonlinearity. The latter step in this work is a simple forward Euler step, which is 1st order in time. It is accomplished by the VQA. In the Methods section we describe in detail the basic idea of the VQA and the pseudospectral split-step method. To this end, we discretize the original wave function $\Psi(x,t)$ on the interval $L$ in space and time, thus obtaining $\Psi_{\rm num}(x_j, t_m)$ with $x_j$ and $t_m$ being integer multiples of a uniform grid spacing $\Delta x$ and an equidistant time step $\Delta t$, respectively. The index $m = 0,...,M-1$ with $M=2^n$ and $n$ the number of qubits. In the quantum algorithm, $\Psi_{\rm num}(t)$ corresponds to an $n$-qubit state vector $|{\bm \psi}(t)\rangle$.    

Using the split-step method has a central advantage over other methods in that it remains stable for long temporal step widths $\Delta t$ due to the absence of the second derivative in $x$. This is important considering the limited number of time steps that can be done with a VQA, partially owed to the accumulating errors of the optimization. An explicit stepping scheme would not be applicable to solve the NLSE by VQA with the same amount of resources, since it requires a smaller $\Delta t$ while the total amount of accurate VQA steps remains fixed. Furthermore, all required steps can be mapped to a quantum circuit.

VQAs are hybrid quantum-classical algorithms where a parameterized cost function $C$ is minimized by an optimizer \cite{Cerezo2021}. This cost function is evaluated by a parametric quantum circuit, which is composed of $n$ qubits and single- and two-qubit gates, the cost minimum search is performed classically in an $N$-dimensional parameter space that contains a parameter vector ${\bm \lambda}$. This parameter vector ${\bm \lambda}$, which consists of the angles of the single-qubit unitary rotation gates ${\bm \lambda} = (\lambda_1, \lambda_2, \dots, \lambda_N)\in \mathbb{R}^N$, is the input to the algorithm. For each time step, the trial solutions for the NLSE $\Psi_{\rm num}(t+\Delta t)$, represented by normalized quantum state vectors $|{\bm \psi}({\bm \lambda})\rangle$ are generated by the parametric quantum circuit
\begin{equation}
|{\bm \psi}({\bm \lambda})\rangle = U({\bm \lambda})|0\rangle^{\otimes n}\,.
\end{equation}
Because the normalization of $\Psi_{\rm num}$ and $|{\bm \psi}\rangle$ differs, they are related by a constant factor, as further explained in the Methods section.

Henceforth, we will use the parameters $a = 2$, $x_0 = -1$, and $v = 10$ for the initial state \eqref{nlse2} obtained from the periodic solution $\Psi_p(x,0)$ in Eq.~\eqref{ana_p}, unless stated otherwise. We apply the quantum algorithm with $n=6$ qubits, a circuit depth of $d=12$ (see Methods for an exact definition), a time step of $\Delta t = 3 \cdot 10^{-3}$, and $\mathrm{ftol} = 10^{-14}$ to compute solutions for the linear case with $s=0$ and the nonlinear case with $s=1$, the latter leading to soliton propagation. The simulation results are shown in Fig.~\ref{fig:lin_and_nonlin}. Both are obtained for a grid resolution of $M=64$ and 6 qubits. The simulation demonstrates that our proposed quantum algorithm describes the time evolution correctly over a long propagation range. The analytical solution is visually identical to the numerical solution, which is shown only for the last time step, to preserve clarity.

\begin{figure}[t]
	\centering
	\includegraphics[width = 0.5\columnwidth]{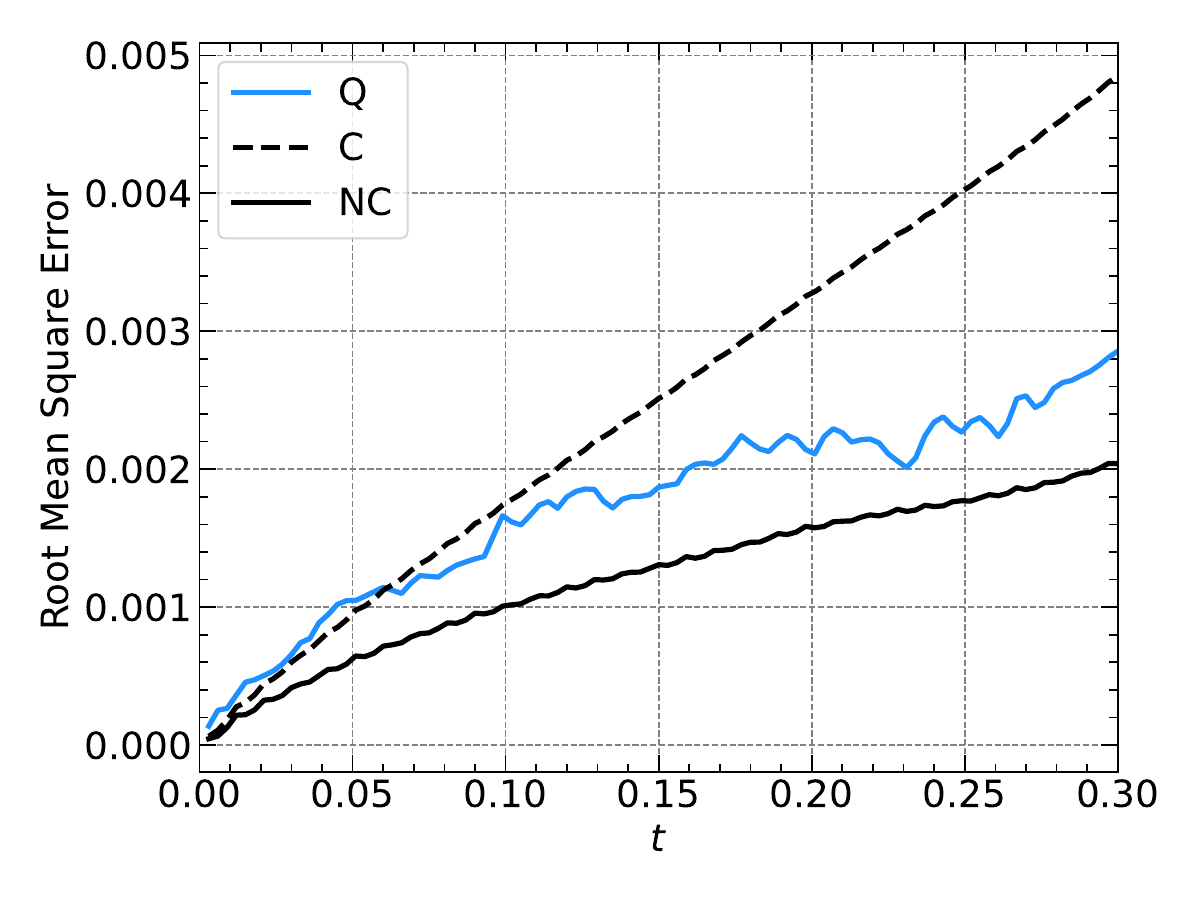}
	\caption{Root mean square (RMS) errors versus time for the simulation shown in Fig.~\ref{fig:lin_and_nonlin}(b) together with the RMS error obtained from the classical versions of the algorithm. (Q) stands for quantum, (C) for classical, and (NC) normalized classical; see the text for further details.}
	\label{fig:nonlin_rms}
\end{figure}

The accuracy of the quantum algorithm will be analyzed in the following by comparing the numerical result for the soliton solution with the analytically exact solution given by Eq.~(\ref{ana_p}). For a quantitative analysis we introduce the root mean square error (RMSE), which is given by
\begin{align}
    \mathrm{RMSE}(m) = \sqrt{\frac{1}{2^n} \sum_{j=0}^{2^n-1} \Big|\,|\Psi_{\rm num}(x_j,t_m)| - |\Psi_p(x_j,t_m)|\,\Big|^2 }\,,
\end{align}
where $\Psi_{\rm num}$ denotes the classical or quantum numerical results, as we detail in the following.

To make a more insightful comparison, we do not only investigate the quantum algorithm (Q), but also the classical version of it (C), as well as the classical version where we manually normalize the solution after each time step (NC), where classical version means that we use the classical algorithm given by Eqs.~(\ref{eq:im_step}) and (\ref{eq:ex_step}). The respective RMSE will be referred to as Q-RMSE, C-RMSE, and NC-RMSE.
Figure~\ref{fig:nonlin_rms} shows all three error measures versus time $t$ for the same parameters as in Fig.~\ref{fig:lin_and_nonlin}.
We see that all errors are on the same order of magnitude. The Q-RMSE is smaller than the C-RMSE for most times $t$, which comes from the intrinsic norm conservation in quantum algorithms, which the classical algorithm lacks. This becomes clear when comparing the Q-RMSE against the NC-RMSE, where we note that the NC-RMSE is a lower bound for the Q-RMSE. Note that due to the random initial conditions before the optimization, the Q-RMSE will depend on the random numbers, but the general trend remains the same. Further simulations show that larger values for $\mathrm{ftol}$ are sufficient, more precisely, a value of $\mathrm{ftol}=10^{-9}$ is sufficient to obtain a Q-RMSE smaller than the C-RMSE for most times.

\begin{figure}[t]
	\centering
	\includegraphics[width = 0.5\columnwidth]{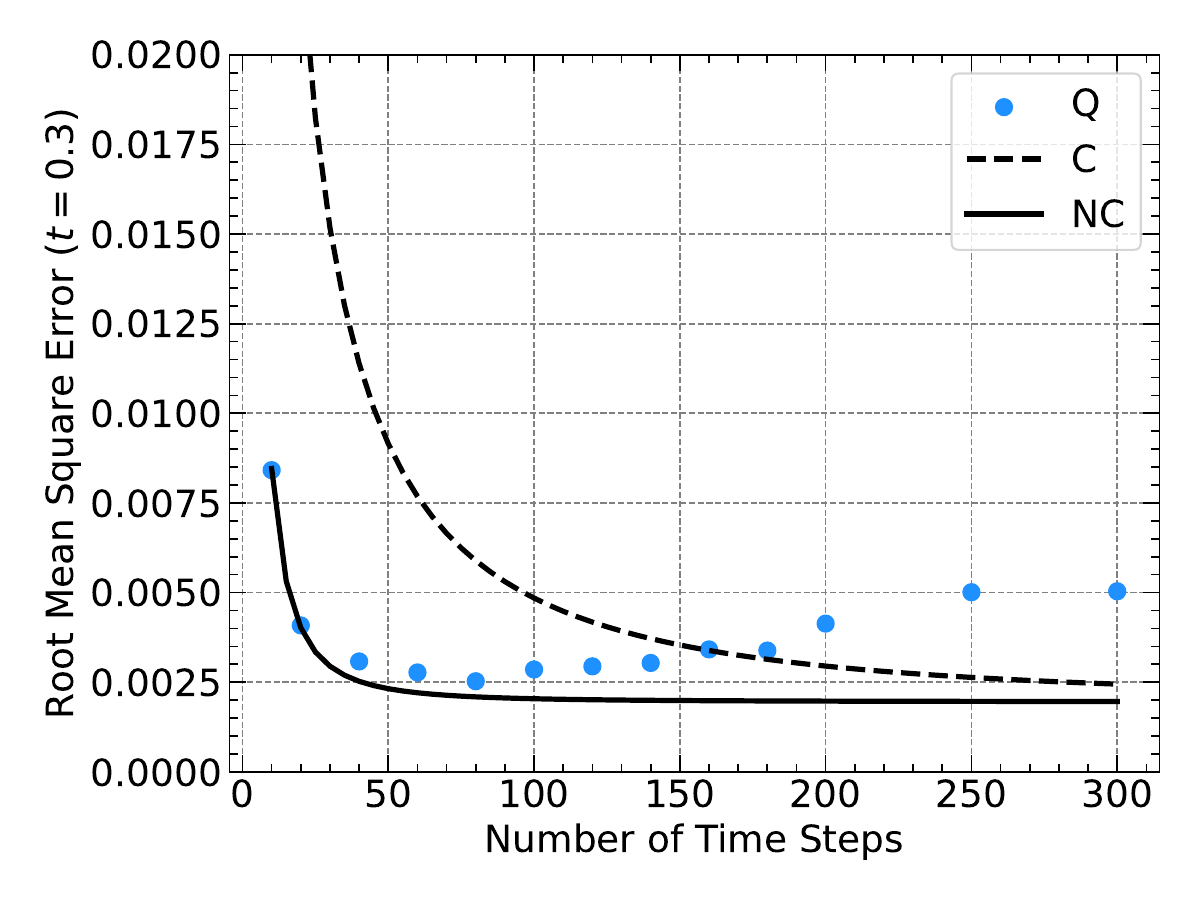}
	\caption{Root mean square error (RMSE) for a VQA simulation with the same parameters as the simulation shown in Fig.~\ref{fig:lin_and_nonlin}(b), but with different numbers of time steps used to span an interval of $t=0.3$ time units. Only the RMS error at $t=0.3$ is shown and is compared with the classical version of the algorithm with and without normalizing the state after every time step. The legend is the same as in Fig. \ref{fig:nonlin_rms}.}
	\label{fig:stepwidth}
\end{figure}

\subsection*{Dependence on Algorithm Parameters}

In this subsection we analyze the impact of parameters, including the timestep width $\Delta t$ and the quantum circuit depth, on the accuracy of the numerical results.

We start by analyzing the dependence on $\Delta t$. Figure~\ref{fig:stepwidth} shows the Q-RMSE for the last simulated time step at $t = 0.3$ for multiple VQA runs, that only differ by the chosen temporal step width $\Delta t$ and thus, in the number of simulated time steps, which are plotted on the $x$-axis. Fig.~\ref{fig:stepwidth} also depicts the C-RMSE and the NC-RMSE. We observe that the Q-RMSE and the NC-RMSE are similar for a wide range of time steps while the latter is always smaller, which is compatible with the result from the previous subsection. For larger numbers of time steps, however, the Q-RMSE exceeds both, the NC-RMSE and the C-RMSE due to the accumulated errors in the classical optimization. For small numbers of time steps, all algorithms produce large RMSE because they become unstable for large $\Delta t$. In conclusion, one can identify a global minimum at which the Q-RMSE is minimized, which corresponds to 80 steps for the fixed integration time interval of $t=0.3$.

\begin{figure}[t]
	\centering
	\includegraphics[width = 0.9\columnwidth]{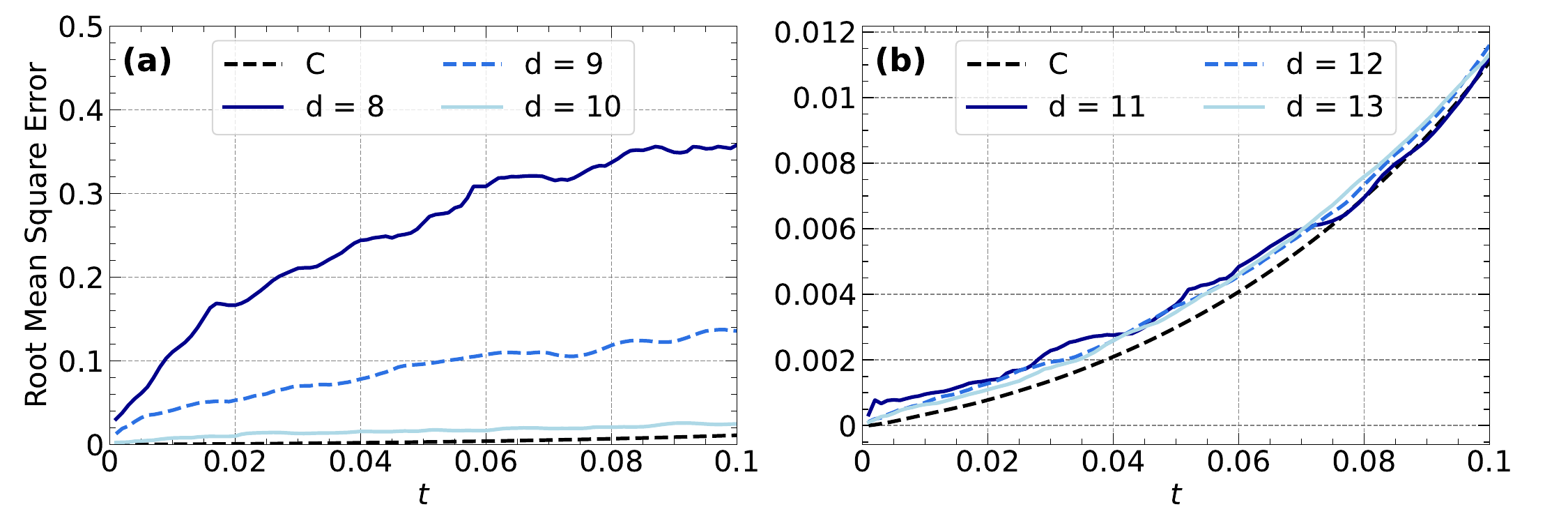}
	\caption{The RMSE is shown for the classical version of the quantum algorithm and for the quantum algorithm with different circuit depths of (a) $d=8$,$9$, and $10$ (b) $d=11$,$12$, and $13$. Qubit number is $n=6$.}
	\label{fig:rms_depth}
\end{figure}

Next, we will discuss the dependence of the Q-RMSE on the circuit depth $d$. The quantum circuit has $2n(d+1)$ free parameters, while the state in Eq.~(\ref{eq:expansion}) has $2^{n+1}$ unknowns, which means that for $n=6$, we have more free parameters than unknowns if $d\ge 10$, i.e., the optimization is overdetermined and underdetermined otherwise. Figure~\ref{fig:rms_depth} shows the Q-RMSE error for 6 VQA runs using $\Delta t = 10^{-3}$, $x_0=0$, and $\mathrm{ftol}=10^{-13}$ together with the respective C-RMSE. The simulations differ only in the depth $d$ of the quantum circuit. We see that the error converges for a depth of $d'=11$, as deeper circuits do not reduce the error significantly. This behavior is expected, since the optimization is overdetermined for $d\ge d'$. In contrast, the Q-RMSE for $d<10$ is large, since the optimization is underdetermined. Overparametrizations have been discussed for quantum neural networks recently \cite{Larocca2023}. In that case, the authors could derive an upper bound on the number of network parameters by means of the dimension of the Lie algebra, the latter of  which is formed by the infinitesimal generators of the network unitary.

\begin{figure}[t]
	\centering
	\includegraphics[width = 0.45\columnwidth]{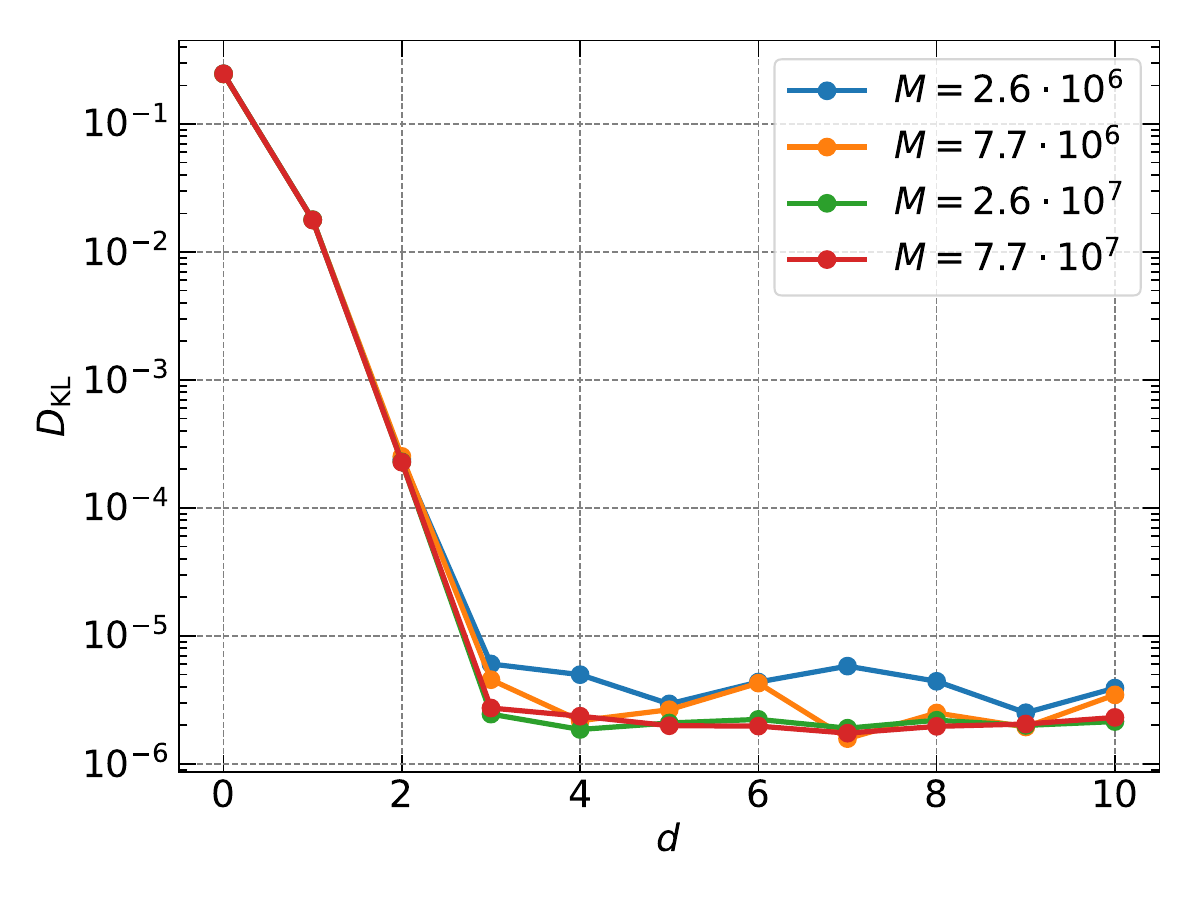}
	\caption{Expressibility of the chosen parametric quantum circuit as a function of the depth $d$ which are explained in the methods section. The legend displays the number of pairs $M$ that have been generated.}
	\label{fig:expr}
\end{figure}

The expressibility of the parametric quantum circuit (PQC) can be quantified using a Kulback-Leibler divergence $D_{\rm KL}$ of two probability density functions (PDF) \cite{Sim2019}, the distribution of fidelities $0\le F\le 1$ of Haar random states $p_{\rm Haar}(F)$ and the distribution of $F$ resulting from the PQC, denoted as $p_{\rm PQC}(F; {\bm \lambda})$. Zyczkowski and Sommers \cite{Zyczkowski2005} obtained $p_{\rm Haar}(F)=(N-1)(1-F)^{N-2}$ with $N=2^n$. Thus follows \begin{equation}
D_{\rm KL}=\int_0^1  p_{\rm PQC}(F; {\bm \lambda})\,\log\left(\frac{p_{\rm PQC}(F; {\bm \lambda})}{p_{\rm Haar}(F)}\right) dF.
\end{equation}
This measure quantifies the ability of a PQC to explore the Hilbert space through an entangled unitary circuit, which is indicated by the closeness of the PDF generated by the PQC with that of the Haar distribution \cite{Sim2019,Holmes2022}. Thus, the smaller the $D_{\rm KL}$, the higher the expressibility. The calculation of $p_{\rm PQC}(F; {\bm \lambda})$ is done as follows: (1) we take 2 random parameter vectors ${\bm \lambda}_1$ and ${\bm \lambda}_2$ and apply the unitary PQC with these input parameter sets to $|0\rangle^{\otimes n}$; (2) this procedure is repeated for $M$ pairs of state vectors to obtain  fidelities $F=|\langle {\bm \psi}({\bm \lambda}_1)|{\bm \psi}({\bm \lambda}_2)\rangle|^2$ and the corresponding PDF. Figure \ref{fig:expr} displays the results for $D_{\rm KL}$ for the 6-qubit PQC as a function of the circuit depth $d$. The number of quantum state pairs is given in the legend. It is seen that the $D_{\rm KL}$ decays rapidly to almost zero (thus maximizing expressibility) for $d\ge 3$ and saturates, similar to Chen et al. \cite{Chen2021} in quantum circuit learning. We have varied the number of pairs $M$ from $2.6 \times 10^6$ up to $7.7 \times 10^7$, which did not change the result. This indicates that the chosen ansatz attains maximum expressibility around $d=3$ for $n=6$. 

Our analysis shows that this measure becomes insensitive for the circuit depths of $d\ge 8$ in the present case, thus indicating a sufficient condition in depth required to attain maximum expressibility. We conclude that, while expressibility can influence the accuracy of the algorithm,  it may however not be a direct consequence. This can be seen by the contrasting magnitude of influence that depth can have on RMSE and $D_{\rm KL}$, as shown here, which is analogous to overfitting. The ability to tune expressibility however may improve the accuracy and scalability of the algorithm, by removing barren plateaus and the corresponding vanishing gradients\cite{Holmes2022}, that are endemic to variational methods as one explores larger $n$. Further insights would require an enhancement of the qubit number and a repetition of the analysis, which is beyond the scope of the present analysis. A further extension can consist of a detailed exploration of different variants of the ansatz circuit, see our configuration in Fig. \ref{fig:ansatz}. For example, other schemes to entangle the qubits are possible and have been for example detailed in ref. \cite{Nakhl2024}. Here, we stick with the present setup.

\subsection*{Complexity estimate of the VQA algorithm}
The complexity of the present algorithm can be estimated by summation of the operations in the algorithm for one integration step. One distinguishes depth complexity $C_D$, the number of successive circuit layers, gate complexity $C_G$, the number of (single qubit) quantum gates, and the query complexity $C_Q$, which is equal to the number of shots on a quantum device run, $N_{\rm s}$. The conservative definition of the time complexity is given by $C_T=C_G \times C_Q$. The estimates are summarized in Table \ref{tab:comlexity}. While gate and depth complexities scale as ${\cal O}(nd)$ for the ansatz circuit and the quantum nonlinear processing unit (QNPU), the complexity of the first step in the split-step method results in complexities of ${\cal O}(n^2)$. The query complexity $C_Q$ corresponds to the number of shots, $N_s$. The measurement noise error is $\epsilon\sim 1/\sqrt{N_s}$ and thus $N_s=\alpha \epsilon^{-2}$. Following ref.\cite{Lubasch2020}, this $\alpha$ depends primarily on the size of the finite grid resolution ($N$) and more particularly on the ratio $N/N_{\rm min}$, where $N_{\rm min}$ (and thus $n_{\rm min}$ qubits) is the minimum grid size needed to resolve the smallest features of the problem at hand. However, noting that the error from finite grid resolution $\epsilon_{g}\sim (\Delta x)^2 \sim N^{-2}\sim 4^{-n}$, i.e., decays exponentially with $n$, it suffices if $N$ is only slightly greater than $N_{\rm min}$. This therefore requires an $\alpha={\cal O}(1)$, thus making the overall $N_{s}\approx \alpha \epsilon^{-2}$, which can be maintained at only nominal counts that do not grow exponentially with $n$.

\begin{table}[b!]
\renewcommand{\arraystretch}{2}
\centering
\begin{tabular}{|l|l|l|l|}
\hline
Algorithm module & $C_D$ & $C_G$ & $C_T$ \\
\hline
Ansatz circuit & $2+(n+2)d \sim {\cal O}(nd)$ & $2n + 3nd \sim {\cal O}(nd)$ &  $C_G \times \alpha \epsilon^{-2}$\\
\hline
QFT$^{\dagger}\, X^{(n-1)} U_{\rm ph} X^{(n-1)}$ QFT & $2\left[\frac{n(n+2)}{2}\right]+2+n^2\sim{\cal O}(n^2)$ & $2\tilde{\alpha} \left[\frac{n(n+2)}{2}\right]+2+n^2\sim{\cal O}(n^2)$& $C_G \times \alpha \epsilon^{-2}$\\
\hline
QNPU incl. Hadamard test & $C_D^{\rm Ansatz}+2n\tilde{\alpha}+2\sim {\cal O}(nd)$ & 
$3C_D^{\rm Ansatz}+2n\tilde{\alpha}+3\sim {\cal O}(nd)$& $C_G \times \alpha \epsilon^{-2}$\\
\hline
\end{tabular}
\caption{Complexity estimates for the major building blocks of the quantum algorithm. We distinguish between depth complexity $C_D$, gate complexity $C_G$, query complexity $C_Q$, and time complexity $C_T$. The orders of magnitude are also given in each entry. The numerical factor $\tilde{\alpha}\sim 10$ denotes the average number of single qubit gates that are necessary to establish 2-qubit or 3-qubit gates, such as CNOT and Toffoli. Here $\alpha=N/N_{\rm min}={\cal O}(1)$ with  $N_{\rm min}$ the minimal required grid size to resolve the structures in the problem at hand well and $\epsilon^{-2}$ the required accuracy from the Monte Carlo sampling of the ancilla qubit to estimate the cost function.}

\label{tab:comlexity}
\end{table}

\section*{Discussion}
In this work, we introduce a quantum algorithm that combines VQA and pseudospectral split-step methods and demonstrate its suitability for solving the NLSE for sufficiently deep quantum circuits for a longer time period in which a soliton solution propagates over a significant distance in space. The split-step method allowed us to keep a first-order time integration scheme for the nonlinear terms of the NLSE while treating the linear part as an integrating factor in combination with Fourier transforms. This keeps the quantum circuit implementation feasible and allowed us to run the algorithm for longer time intervals than typically seen.

Variational quantum algorithms rely on the minimization of a cost function in a high-dimensional parameter space. This is done in the classical part of the hybrid algorithm. A non-monotonicity of the root mean square error (RMSE) versus time can result from such an optimization procedure, which was discussed for example in ref. \cite{Ingelmann2024}. Such behavior is also observed in the present work, see e.g. Fig. \ref{fig:nonlin_rms}. This can become a critical point for a VQA since the minimization of $C({\bm \lambda})$ is typically a non-convex optimization problem. In the absence of an analytical or a classical numerical solution, a convergence criterion for the VQA is then difficult to formulate. To the best of our knowledge, a solution to this problem can then only lie in either the comparison of different classical optimization methods for the same problem or the construction of surrogate-based optimization steps using local kernel-based approximations as suggested in ref. \cite{PhysRevA.107.032415}. The required classical optimization also restrains the scalability for larger systems with more qubits by requiring a high-dimensional classical optimization, which constitutes another challenge for this solution method. However, besides these potential limitations, such as  (i) barren plateaus, (ii) the scalability of the high-dimensional optimization problem and (iii) the difficulty of finding a convergence criterion, our results demonstrate that the algorithm works robustly and generates accurate results.

The present work should be considered as a proof-of-concept study. A continuation of the study along these lines should imply possible future steps: (i) a switch to a full quantum simulation with realistic shot noise and gate errors and, eventually (ii) an implementation on a NISQ device. Realizing this would typically be faced with practical bottlenecks including decoherence of qubits, hardware and shot noise. Particularly, the interplay between noise and barren-plateaus, which is endemic to VQA methods \cite{wang2021,liu2025}, determines the overall performance. The presence of noise can either worsen \cite{wang2021} or in certain cases can also alleviate \cite{liu2025} the problem of barren-plateaus, which is determined by the type and magnitude of noise present. The specifics of these steps will be reported elsewhere.

\section*{Methods}\label{sec:methods}

\subsection*{Pseudospectral Split-Step Method}

We consider the solution on a finite domain $x \in [-\pi,\pi)$ and discretize the interval with $M$ grid points $x_j = -\pi + 2\pi j/M$ ($0\le j < M$). We use periodic boundary conditions such that $x_M \equiv x_0$. We introduce the linear operator $\mathcal{L}$ and the nonlinear operator $\mathcal{N}$ as
\begin{align}
    \mathcal{L}\psi = -\frac{1}{2} \partial_{xx} \psi\quad\text{and}\quad \mathcal{N}\psi = -s|\psi|^2 \psi,
\end{align}
such that we can rewrite the NLSE \eqref{nlse1} as 
\begin{align}
    \partial_t \psi = -i(\mathcal{L} + \mathcal{N})\psi,
\end{align}
with formal solution at $t+\Delta t$
\begin{align}
    \psi(x,t+\Delta t) = \exp(-i\Delta t(\mathcal{L}+\mathcal{N})) \psi(x,t).
\end{align}
We apply the Baker-Campbell-Hausdorff formula \cite{Hall2015}
\begin{align}
     \exp(-i\Delta t \mathcal{N})\exp(-i\Delta t \mathcal{L}) = \exp\left(-i\Delta t(\mathcal{L}+\mathcal{N}) - \frac{\Delta t^2}{2}[\mathcal{N},\mathcal{L}] + \mathcal{O}(\Delta t^3)\right),
\end{align}
and neglect all terms of order $\mathcal{O}(\Delta t^2)$, such that the operators can be applied separately in a two-step process that is first-order correct \cite{pathria1990pseudo,fioroni2024python}. For the solution of the linear Schrödinger equation with different potentials by a spectral method we also refer to ref. \cite{Feit1982}. First, we compute an implicit substep for the linear Laplacian operator using the exact solution in Fourier space, which implicitly includes the periodic boundary condition. The wavenumbers of the Fourier modes are given by $k_j=j-M/2$ ($0\le j < M$). Then, we compute an explicit step for the nonlinear operator by using an Euler step. The time-discretized stepping scheme from $t$ to $t+\Delta t$ proceeds in two substeps which take the following form in a classical notation. The first implicit substep is given by \cite{Benenti2008}
\begin{align}
\mbox{Step 1:} \quad \tilde{\Psi}_{\rm num}(x_j,t)
= \exp\bigl(-i V \Delta t\bigr) \mathcal{F}^{-1}\left( \exp\left(-\frac{i}{2} k_j^2 \Delta t\right) \mathcal{F}\left(\Psi_{\rm num}(x_j,t)\right)\right)\,.\label{eq:im_step}
\end{align}
For the case of $V=0$ examined here, the exponential term $\exp(-i V \Delta t)$ is equal to the identity. Subsequently, the second explicit (Euler) substep follows to
\begin{align}
\mbox{Step 2:} \quad \Psi_{\rm num}(x_j,t+\Delta t) = \tilde{\Psi}_{\rm num}(x_j,t) + i s \Delta t |\tilde{\Psi}_{\rm num}(x_j,t)|^2 \tilde{\Psi}_{\rm num}(x_j,t)\,.\label{eq:ex_step}
\end{align}
The symbols $\mathcal{F}$ and $\mathcal{F}^{-1}$ stand for the discrete Fourier and inverse Fourier transforms, respectively.

In the next Methods subsection, we discuss the realization of this scheme with a quantum algorithm. Note that the implicit step Eq.~(\ref{eq:im_step}) requires two Fourier transforms and a multiplication by a phase. The latter can be done with a circuit shown in ref.  \cite{Benenti2008}, while the former is realized by a straightforward application of the quantum Fourier transform algorithm \cite{Nielsen2010,wright2024noisy}. The explicit step can be computed by VQA.

\subsection*{Evaluation of the Cost Function and Circuit Implementation}\label{sec:methods_d}
The cost function $C({\bm \lambda})$ is then also evaluated on the quantum device in a Hadamard test-like circuit \cite{Lubasch2020,Ingelmann2024}. In detail, the cost function for the solution of a PDE is given by
\begin{align}
C({\bm \lambda}) = \| |{\bm \psi}({\bm \lambda})\rangle - F[|{\bm \psi}(t)\rangle]\|^2_2
= C_1-2|\langle {\bm \psi}({\bm \lambda}) |F[{\bm \psi}(t)\rangle]|^2\,,\label{eq:cost_function}
\end{align}
where $F[|{\bm \psi}(t)\rangle]$ is the (nonlinear) iteration from the past step $t$ in correspondence with the underlying partial differential equation (PDE) and $C_1$ is a constant. The cost function is minimized by maximizing the scalar product (or overlap) in the second term of eq. \eqref{eq:cost_function}. Note, that one can drop the constant term in the cost function for the evaluation. Multiple repeated measurements of identically prepared quantum states per iteration step evaluate the costs. These costs are minimized with a limited-memory Broyden-Fletcher-Goldfarb-Shanno algorithm with constant lower and upper bounds for $\lambda_k$ (L-BFGS-B) which applies a quasi-Newton method for solving unconstrained, nonlinear optimization problems \cite{Byrd1995}. Thereby, the Hessian matrix of the cost function is approximated by the evaluation of the gradients (or the approximated gradients) in order to find the descent direction in the hyperparameter landscape. The optimal parameter set ${\bm \lambda}^*$ initializes the ansatz function such that the solution of the given problem can be observed \cite{Cerezo2021}. We use random initial parameters for the first time step and the optimized parameters from the previous step for all subsequent steps.

We expand the state vector into the $n$-qubit basis, using the complex wave function as expansion coefficients
\begin{align}
    |{\bm \psi}(t)\rangle = \sum_{j=0}^{2^n-1} \psi_{j}(t)|j\rangle = \sqrt{\frac{\Delta x}{2 a}}\sum_{j=0}^{2^n-1} \Psi_{\rm num}(x_j,t)|j\rangle \label{eq:expansion}
\end{align}
where $|j\rangle$ denotes the $n$-qubit basis state corresponding to the binary representation of $j$ with expansion coefficients $\psi_j(t)= \sqrt{\Delta x/(2 a)}\,\Psi_{\rm num}(x_j,t)$. Consequently, $M=2^n$ with a qubit number $n$. Equation~(\ref{eq:expansion}) provides a construction for switching between the representations and satisfies the normalization constraints
\begin{align}
    \langle {\bm \psi}(t) | {\bm \psi}(t)\rangle &= 1, \label{eq:norm1}
\end{align}
and
\begin{align}
\int_{-\infty}^{\infty} |\Psi(x,t)| dx &= 2 a. \label{eq:norm2}
\end{align}
Due to Eq.~(\ref{eq:norm2}), we do not need an extra optimization parameter for the amplitude. The computed solution is related to the solution of the NLSE by a constant factor. Note that the initial condition for the soliton at $t=0$ is directly implemented in the cost function.

We formulate a cost function for the operator $F$ that describes the explicit step Eq.~(\ref{eq:ex_step}). $F$ is diagonal and its elements $f_{j,j}$ have the form:
\begin{align}
    f_{j,j} = 1 + i s \Delta t \frac{2a}{\Delta x} |\psi_j(t)|^2 \,.
\end{align}
Note that the factor $2a/\Delta x$ originates from the normalization conditions Eqs.~(\ref{eq:norm1}) and (\ref{eq:norm2}).

Substituting $F$ into Eq.~(\ref{eq:cost_function}) and omitting constant terms yields the following cost function 
\begin{align}\label{eq:costFunction}
    C({\bm \lambda}_{t+\Delta t}) = s \Delta t \frac{2 a}{\Delta x} \mathrm{Im}\left\{\langle {\bm \psi}(t+\Delta t)| \tilde{\Psi}_{\rm num}(t) \tilde{\Psi}_{\rm num}^*(t) | \tilde{\bm \psi}(t) \rangle \right\} 
    -\mathrm{Re}\left\{\langle {\bm \psi}(t+\Delta t) | \tilde{\bm \psi}(t) \rangle \right\},
\end{align}
where $|\tilde{\bm \psi}(t)\rangle$ is the quantum state vector after the implicit substep and $|{\bm \psi}(t+\Delta t) \rangle$ is a function of ${\bm \lambda}_{t+\Delta t}$.

\begin{figure}[t]
    \centering
    \includegraphics[width=0.6\columnwidth]{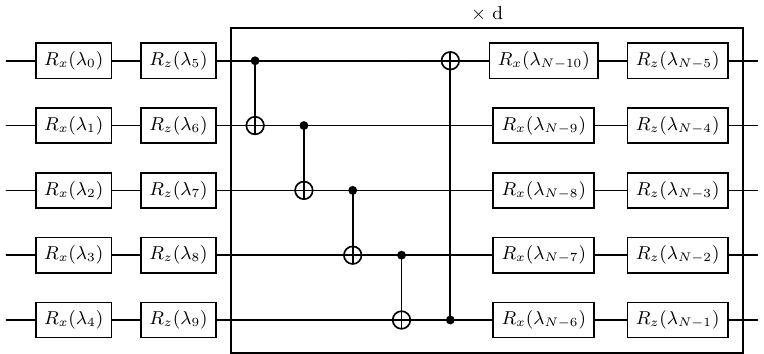}
    \caption{Ansatz circuit for $n=5$ qubits. The depth $d$ determines the number of layers of controlled NOT (CNOT) gates, each followed by single qubit rotations $R_x$ and $R_z$ that are applied after an initial layer of rotation gates.}
    \label{fig:ansatz}
\end{figure}
\begin{figure}[t]
    \centering
    \includegraphics[width=0.6\columnwidth]{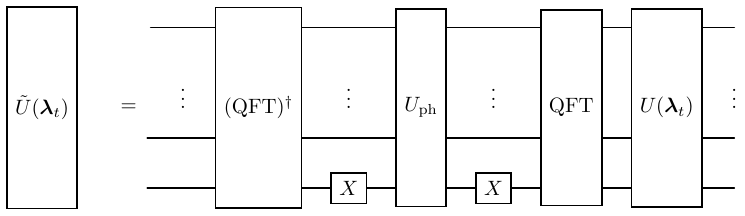}
    \caption{Circuit for initializing the trial state and computing the first substep of the split-step method.}
    \label{fig:U_tilde}
\end{figure}

To initialize the state $|{\bm \psi}(t)\rangle = U({\bm\lambda}_t)|0\rangle^{\otimes n}$ and the trial state $|{\bm \psi}(t+\Delta t)\rangle = U({\bm\lambda}_{t+\Delta t})|0\rangle^{\otimes n}$ we use the parameterized ansatz circuit shown in Fig.~\ref{fig:ansatz} consisting of multiple layers of single-qubit rotations and controlled NOT gates, where we define the depth $d$ of the quantum circuit as the number of layers of controlled NOT (CNOT) gates, each followed by a layer of single qubit rotation gates $R_x$ and $R_z$, that are applied after the initial layer of rotation gates. Note that we tested other ansatz circuits as well, however, the one presented here was the most accurate one. For each time step, the state $|\tilde{\bm \psi}(t)\rangle$ after the first substep is then obtained by additionally applying the quantum Fourier transform (QFT) \cite{Nielsen2010}, multiplying by the appropriate phase for each $k$ in Fourier space, see Eq.~\eqref{eq:im_step}, and applying the inverse QFT. This includes 
\begin{align}
    | \tilde{\bm \psi}(t) \rangle = \tilde{U}({\bm\lambda}_t) |0\rangle^{\otimes n} \,,
\end{align}
with
\begin{align}
\tilde{U}({\bm\lambda}_t)=({\rm QFT})^\dagger X^{(n-1)} U_{\rm ph} X^{(n-1)} ({\rm QFT}) U({\bm\lambda}_t)\,.
\end{align}
After applying the QFT, the zero frequency component of the state is shifted to the center of Fourier space by applying an $X$ gate to the most significant qubit and another $X$ gate before the inverse QFT reverses this shift. Fig.~\ref{fig:U_tilde} shows a diagram of the circuit corresponding to this substep. Following reference \cite{Benenti2008}, $U_{\rm ph}$ can be constructed using $n^2$ phase and controlled phase gates
\begin{equation}
    U_{\rm ph} = \prod_{i,j=0}^{n-1} P_{ij}
\end{equation}
with 
\begin{equation}
    P_{ij} = 
        \begin{cases}
            P^{(i)}(\gamma (2^{2i}-2^{n+i})) & \text{if } i=j \\
            CP^{(ij)}(\gamma 2^{i+j}) & \text{if } i \neq j
        \end{cases}
\end{equation}
where $\gamma = -\Delta t/2$. $P$ and $CP$ denote the phase and controlled phase gate, respectively.

\begin{figure}[t]
    \centering
    \includegraphics[width=0.6\columnwidth]{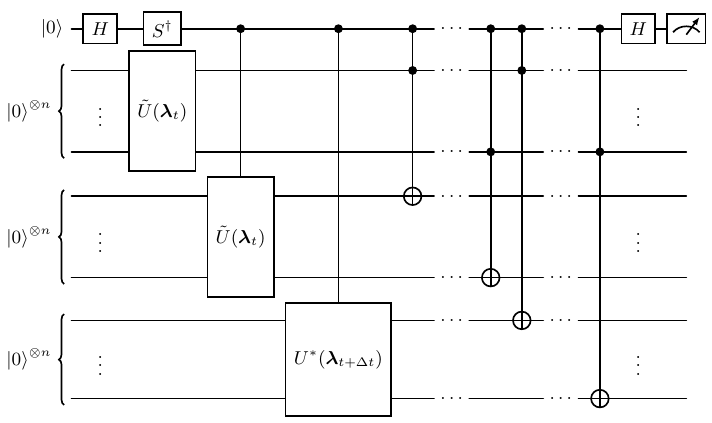}
    \caption{Circuit for computing the nonlinear term in the cost function. Its value is obtained via a computational basis measurement of the ancilla qubit.}
    \label{fig:nonlinearCost}
\end{figure}

Using these circuits the second term in Eq.~(\ref{eq:costFunction}) can be computed using a Hadamard test, see e.g. ref.~\cite{Ingelmann2024} for a detailed description. Figure \ref{fig:nonlinearCost} shows the circuit used to evaluate the nonlinearity, denoted as the quantum nonlinear processing unit (QNPU), which evaluates
\begin{equation}\label{eq:nonlinearTerm}
  \textrm{QNPU}= \mathrm{Im}\left\{ \sum_{j=0}^{2^n-1} \psi_j^*(t+\Delta t) |\tilde{\psi_j}(t)|^2 \tilde{\psi_j}(t) \right\}\,.
\end{equation}
The QNPU is based on the circuits from refs. \cite{Lubasch2020,Jaksch2023} for evaluating functions ${\cal F}$ of the form
\begin{equation}
    {\cal F} = f^{{(1)}^*} \prod_{j=1}^r f^{(j)}.
\end{equation}
To arrive at the circuit shown here we use $r=3$, $f^{(1)}=f^{(2)}=\tilde{\psi}(t)$ and $f^{(3)}=\psi^*(t+\Delta t)$ and add an $S^\dagger$-gate to compute the imaginary part instead of the real part of the expression. The value of expression~(\ref{eq:nonlinearTerm}) is then obtained as the expectation value of a Z-measurement performed on the ancilla (first) qubit. The unitary $U^*({\bm\lambda}_{t+\Delta t})$ initializes the complex conjugate of the state $|{\bm \psi}(t+\Delta t)\rangle$ and is obtained by applying $U({\bm\lambda}_{t+\Delta t})$ but changing the sign of the angles of rotation of the $R_x$ and $R_z$ gates.

To reduce the computation time for evaluating the cost function, we do not directly compute the required expectation values by performing full quantum simulations of the circuits presented in this section. We instead obtain the trial state prepared by the Ansatz circuit using the quantum simulation software Qiskit \cite{Qiskit} and from this directly compute the value of the cost function. The QFT algorithm is a well-established building block of quantum algorithms and it would be straightforward to include in the present scheme. This has been done in for example in the variational algorithm of Huang and co-workers\cite{Huang2023}. We verified that the results obtained with our approach match the results obtained with ideal quantum simulations of the presented circuits. The classical optimization of the quantum circuit is done with the Qiskit implementation of the L-BFGS-B optimizer algorithm. A relevant parameter for the optimization algorithm is $\mathrm{ftol}$ which is an upper bound for the relative error of two consecutive iteration steps that must be satisfied for the optimization to terminate.

Qiskit can be used in three different ways, (1) as an ideal simulator without quantum circuit noise monitoring the complex quantum statevector (statevector simulator), (2) as an ideal simulator without quantum circuit noise and with measurement noise (qasm simulator), and (3) as a simulator that emulates the NISQ devices with quantum circuit and measurement noise. We will only use (1) throughout this work.

\bibliography{apssamp}

\section*{Data availability}
A repository containing an example script of our quantum simulation program for the NLSE, which was used for the present work, can be accessed via ref.\cite{kocher_2024_13986469}. The datasets generated for the current study are available from the corresponding author on reasonable request.

\section*{Acknowledgements}
The authors gratefully acknowledge the computing time made available to them on the high-performance computer Noctua 2 at the NHR Center Paderborn Center for Parallel Computing (PC$^2$). This center is jointly supported by the Federal Ministry of Education and Research and the state governments participating in the NHR (www.nhr-verein.de/unsere-partner). The work of J.S. is supported by the European Union (ERC, MesoComp, 101052786). Views and opinions expressed are however those of the author only and do not necessarily reflect those of the European Union or the European Research Council. Neither the European Union nor the granting authority can be held responsible for them. We acknowledge support for the publication costs by the Open Access Publication Fund of the Technische Universität Ilmenau. 

\section*{Author contributions statement}

N.K., H.R., J.S. and S.S. designed the research. N.K. and H.R. conducted the simulations. S.S.B. conducted the complexity analysis of the algorithm. All authors analysed and discussed the simulation results and wrote the manuscript. 

\section*{Competing interests}
The authors report no conflicts of interest.

\end{document}